# BARYON-BARYON POTENTIALS ON THE LATTICE [1]

K. RABITSCH, H. MARKUM, W. SAKULER

*Institut für Kernphysik, Technische Universität Wien*
*A-1040 Vienna, Austria*

The interaction of spatially extended heavy baryons is investigated in the framework of lattice QCD with dynamical quarks. It is shown that the expected dipole forces have a very short range and that the baryon-antibaryon interaction is more attractive than the baryon-baryon interaction. Sea quarks play a minor important role.

---
[1]Supported in part by "Fonds zur Förderung der wissenschaftlichen Forschung" under Contract No. P7510-TEC.

*1. Introduction.* Quark theory provides new degrees of freedom at the subnuclear level which were previously considered by meson theory. The vacuum of QCD contains both virtual gluons and quarks. As a consequence the nucleon-nucleon forces are mediated for short distances by gluon exchange between the constituent quarks whereas for longer distances the production of quark-antiquark pairs is the dominating mechanism. The quark-antiquark exchange can be treated as an effective meson exchange leading to the construction of the Bonn and Paris potentials [1]. The meson theoretical potentials give a satisfactory description of the nucleon-nucleon scattering data which are mainly sensitive to long range distances. The gluon exchange can be studied by phenomenological potential and bag models allowing for a first insight into the interaction mechanism of the six-quark system [2].

Both quark and meson potentials contain parameters and are based on phenomenology. Today the aim should be to calculate the nucleon-nucleon forces directly from the field equations of QCD. Although we are at present far from a treatment of complete QCD from first principles some attempts can be performed in the framework of a restricted QCD. In the last decade the simulation of quantum field theories on computers made fast progress and reached a high standard. Thus, one should start now with the investigation of nucleon-nucleon interactions on the basis of a truncated (and maybe unrealistic) QCD. One restriction is to neglect quark loops in the action and to take only gluon exchange into account. Another approximation is to reduce the dynamical propagators of the valence quarks of the hadron to static propagators. This can be weakened by making one propagator dynamical and leaving the others as spectator quarks. With the development of the next computer generations the codes can be extended step by step in order to take finally the full QCD process into account.

During the last years lattice calculations have demonstrated that the potential between a static quark and a static antiquark of a meson is confining. The same result turns out for the three quarks of a baryon [3]. All these outcomes of the static approximation make us believe QCD to be the proper theory of strong interactions. Thus, an application of static QCD to the nucleon-nucleon system seems interesting. First investigations of the gluon exchange between two three-quark clusters yielded an attractive potential limited to the overlap region of the baryons [4]. The extension of the pure gluon Lagrangian to the full QCD Lagrangian provides for the creation of virtual quark-antiquark pairs out of the vacuum. Many conceptual and technical problems have been overcome and calculations with dynamical fermions are now feasible [5]. Thus, in this paper we simulate the meson exchange between two static three-quark clusters induced by virtual quark-antiquark pairs and calculate the potential energy with varying distance of the two nucleons. By studying such quark-antiquark pairs representing virtual mesons we gain some insight into QCD based meson exchange.

*2. Theory.* In our approximation the valence quarks of a hadron are restricted to fulfill the static Dirac equation. The gluons are treated as SU(3)-Maxwell fields $U_{x\mu}$ and the sea quarks as Dirac fields $\bar{\psi}_x, \psi_x$ in Kogut-Susskind discretization. The free energy $F_N(\vec{r}_1, \ldots, \vec{r}_N)$ of a system of $N$ quarks and antiquarks in a gluonic and fermionic field is defined by the thermodynamical expectation value [6]

$$\exp\left(-\frac{1}{T}F_N(\vec{r}_1, \ldots, \vec{r}_N)\right) := \frac{1}{3^N} \sum_{|s_q s_g s_f\rangle} \langle s_{q_1} \ldots s_{q_N} s_g s_f | e^{-\frac{1}{T}H} | s_{q_1} \ldots s_{q_N} s_g s_f \rangle . \quad (1)$$

The Hamiltonian $H$ contains the dynamics of the system and the temperature $T$ is equivalent to the inverse Euclidean time extension $N_t a$ with lattice unit $a$. The expectation value has to be taken over all static quark states $|s_{q_1} \ldots s_{q_N}\rangle$ and over all states $|s_g s_f\rangle$ of the gluon and fermion fields. Writing the time evolution $e^{-Ht}$ of the quark states $|s_q\rangle$ in terms of the static quark propagator, the so-called Polyakov loop

$$L(\vec{r}) = \frac{1}{3} \operatorname{Tr} \prod_{i=1}^{N_t} U_{\mu=4}(\vec{r}, t_i) , \quad (2)$$

leads to the expression

$$\exp\left(-\frac{1}{T}F_N(\vec{r}_1, \ldots, \vec{r}_N)\right) = \sum_{|s_g s_f\rangle} \langle s_g s_f | e^{-\frac{1}{T}H} L(\vec{r}_1) \ldots L(\vec{r}_N) | s_g s_f \rangle . \quad (3)$$

This expectation value in the Euclidean Hamiltonian formulation can be recasted into a path integral in the Lagrange form by identifying the states $|s_g s_f\rangle$ with the field variables $U, \bar{\psi}, \psi$. Normalization with respect to the quark-vacuum energy $F_0$ yields the Feynman path-integral representation for the expectation value of a product of $N$ Polyakov loops

$$\exp\left(-\frac{1}{T}F(\vec{r}_1,\ldots,\vec{r}_N)\right) = \frac{\int \mathcal{D}[U]\mathcal{D}[\bar{\psi},\psi]\, L(\vec{r}_1)\ldots L(\vec{r}_N)\, e^{-S[U,\bar{\psi},\psi]}}{\int \mathcal{D}[U]\mathcal{D}[\bar{\psi},\psi]e^{-S[U,\bar{\psi},\psi]}}$$
$$= \langle L(\vec{r}_1)\ldots L(\vec{r}_N)\rangle. \tag{4}$$

$F(\vec{r}_1,\ldots,\vec{r}_N) := F_N(\vec{r}_1,\ldots,\vec{r}_N) - F_0$ is the net free energy of the $N$-quark system and $S(U,\bar{\psi},\psi) = S_G(U) + S_F(U,\bar{\psi},\psi)$ the total action of the system. For the gluonic action $S_G$ we use Wilson's plaquette action

$$S_G = \beta \sum_{x\mu}(1 - \frac{1}{3}\,\mathrm{Re\,Tr}\, U_{x\mu}U_{x+\hat{\mu},\nu}U^{\dagger}_{x+\hat{\nu},\mu}U^{\dagger}_{x\nu}), \tag{5}$$

with the inverse gluon coupling $\beta = 6/g^2$, and for the fermionic action $S_F$ we employ the Kogut-Susskind formulation [7]

$$S_F = \frac{n_f}{4}\left\{\sum_{x\mu}\Gamma_{x\mu}\frac{1}{2}(\bar{\psi}_x U_{x\mu}\psi_{x+\hat{\mu}} - \bar{\psi}_{x+\hat{\mu}}U^{\dagger}_{x\mu}\psi_x) + m\sum_x \bar{\psi}_x\psi_x\right\}, \tag{6}$$

where $n_f$ is the number of flavors and the $\Gamma_{x\mu}$ play the role of the Dirac matrices.

In order to compute the path integral (4) we formulate the theory on a space-time lattice and perform Monte Carlo simulations. We take a lattice of size $8 \times 8 \times 16 \times 4$ with periodic boundary conditions. We compare QCD with and without dynamical fermions applying the pseudofermionic method. In the pure gluonic case we choose $\beta = 5.6$ which corresponds via the renormalization group equation to a lattice constant of $a \approx 0.25$ fm and to an effective spatial extension in $z$ direction of about 2 fm. For full QCD we set the number of flavors to $n_f = 3$ at $\beta = 5.2$ and to $n_f = 4$ at $\beta = 4.9$ with the dynamical quark mass $m = 0.1$. This choice leads to comparable lattice constants in the confinement regime.

We construct a quark wave function $\Psi(\vec{r})$ by distributing the static quark $L(\vec{r})$ in Gaussians

$$\Psi(\vec{r}) = (\sqrt{\pi}\sigma)^{-3/2}\, e^{-\frac{r^2}{2\sigma^2}} \tag{7}$$

over a sphere with radius $R$ on the lattice. Since the results turn out to depend only weakly on the width $\sigma$ we set (7) to a uniform distribution. For the description of baryons we take the product of three quark wave functions [8]. Because the static quarks carry no spin and flavor all hadrons are degenerate. To study the gluon and quark-antiquark exchange between baryons the total wave function of the baryon-baryon system is constructed to be a product of Gauss functions with two centers separated by some distance $d$.

*3. Results.* We calculate the total energy of the hadron-hadron system according to (4) and subtract twice the energy of a single baryon. In Fig. 1a we display the relative potential energy of the two-baryon system with baryons of radius $R = 1a$ for varying distances $d$ between the two clusters. We find an attractive potential with an effective range of about $2R$. The reason for the attractive forces between the two baryons lies in the Coulomb plus linear type of potential between quarks which tries to attract the two three-quark clusters. When the two baryons become separated their colors are saturated and they evolve to isolated color singlets which do not interact. At least on the lattice no long range forces can be resolved. Taking dynamical fermions into account we obtain no drastic effect. This is also observed in mass spectrum calculations where sea quarks give rise to only a 10 per cent effect. In Fig. 1b we display the potential energy for a two-baryon system with baryon radius $R = 2a$. We find attractive interactions only in the overlap region and practically no flavor dependence.

We now turn to the baryon-antibaryon system in Fig. 1c and again see an attractive interaction in the overlap regime. The difference between the baryon-baryon and baryon-antibaryon potentials can be interpreted qualitatively by the fact that the baryon-antibaryon system can build three mesons which is energetically favored. Switching on dynamical fermions we observe practically no effect. Increasing the radius of the hadron sphere to $R = 2a$ we recognize in Fig. 1d more or less the same behavior as for $R = 1a$.

*4. Conclusion.* Let us discuss the main deficiencies of our approach and propose some improvements for future more realistic studies of nucleon-nucleon interactions. We work with a truncated Dirac equation restricting the valence quarks to static color charges.

The calculated energy of the considered quark wave functions has thus the meaning of the energy content of quark charge distributions in a gluon field. This is equivalent to the potential of an electric charge distribution in electrostatics. It must be emphasized that the construction of the quark orbitals leads to baryons of reasonable size but takes no kinetic energy into account.

The ideal would be to evaluate the $S$ matrix for scattering of two nucleons consisting of three quarks each. This necessitates the computation of a 4-point hadron-Green-function which is equivalent to a correlation function of six quark propagators in the QCD path-integral. A feasible approximation might be to make two quarks of a baryon heavy so that they only act as spectator quarks in the scattering process [9]. There has been some progress in computing meson scattering-amplitudes from the finite volume dependence of the bound-state energies [10, 11]. A method to extract effective potentials from the two-hadron spectrum was proposed recently [12].

This paper contains a first study of the influence of both gluons and dynamical quarks on static baryon-baryon systems. The sea quarks constitute the meson exchange as expected by the Yukawa theory. With the extension of the pure gluon Lagrangian to the full QCD Lagrangian the contribution of virtual mesons was investigated. It turned out that the gluon exchange dominates and sea quark effects are small. We have seen that overlapping baryons feel the colors of the constituent quarks and the resulting confinement potential while two separated three-quark clusters form color singlets which do not interact. We conclude that the residual color forces between color-neutral quark-clusters have a very short range and that the interaction takes place practically only in the overlap region.

**Baryon-Baryon Potentials**

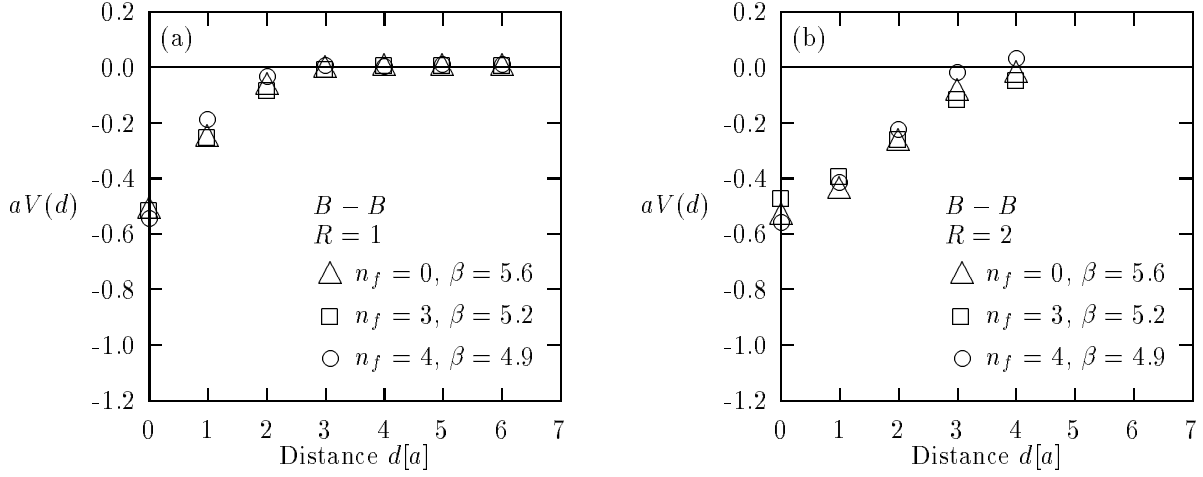

**Baryon-Antibaryon Potentials**

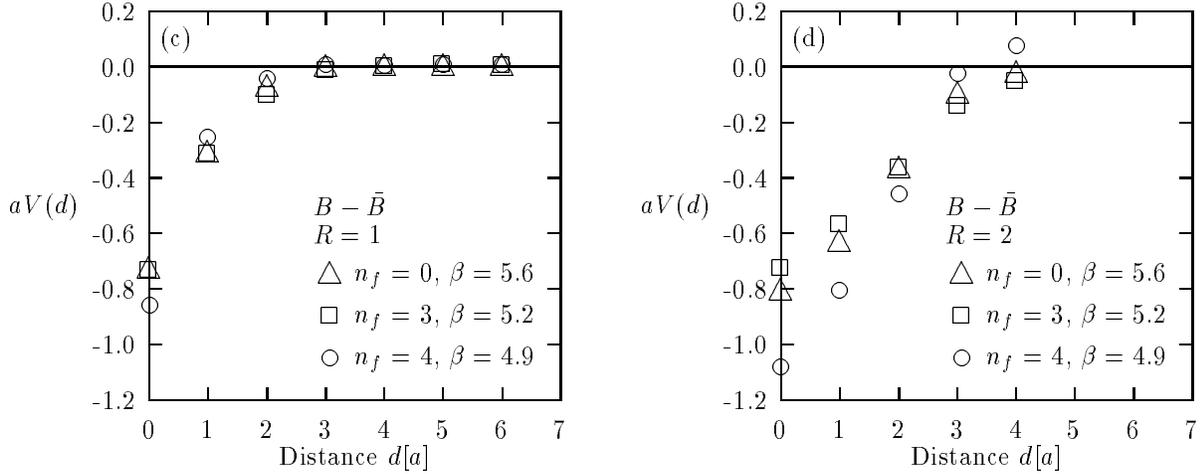

Fig. 1: Baryon-baryon potentials as a function of center of mass distance $d$ for pure gluon exchange and with sea quarks for baryons with radii $R = 1$ and $R = 2$. Comparison of baryon-antibaryon potentials with and without dynamical fermions. The interaction is attractive and takes place mainly in the overlap region. The antibaryon potentials turn out to be deeper due to annihilation into three mesons. Dynamical quarks have no pronounced effect. Error bars are in the size of the symbols.